\documentclass[prd,aps,groupedaddress,showpacs,amsmath]{revtex4}
\usepackage{epsf,epsfig,graphicx}

\newcounter{muni}

\begin{document}
\hbadness=10000 \pagenumbering{arabic}

\preprint{ \vbox{
\hbox{MIT-CTP-3902}   }
}
\title{Possible complex annihilation and $B\to K\pi$ direct CP asymmetry}

\author{Junegone Chay$^{1}$}
\email{chay@korea.ac.kr}
\author{Hsiang-nan Li$^{2}$}
\email{hnli@phys.sinica.edu.tw}
\author{Satoshi Mishima$^3$}
\email{mishima@ias.edu}

\affiliation{$^{1}$Center for Theoretical Physics, Laboratory for
  Nuclear Science, Massachusetts Institute of Technology, Cambridge,
  MA 02139, U.S.A.}
\affiliation{Department of Physics, Korea University, Seoul
  136-701, Korea}
\affiliation{$^{2}$Institute of Physics, Academia Sinica, Taipei,
Taiwan 115, Republic of China} \affiliation{Department of Physics,
National Cheng-Kung University, Tainan, Taiwan 701, Republic of
China} \affiliation{Department of Physics, National Tsing-Hua
University,   Hsinchu, Taiwan 300, Republic of China}
\affiliation{$^{3}$School of Natural Sciences, Institute for
Advanced Study, Princeton, NJ 08540, U.S.A.}

\begin{abstract}

We point out that a sizable strong phase could be generated from the
penguin annihilation in the soft-collinear effective theory for $B$
meson decays. Keeping a small scale suppressed by $O(\Lambda/m_b)$,
$\Lambda$ being a hadronic scale and $m_b$ the $b$ quark mass, in
the denominators of internal particle propagators without expansion,
the resultant strong phase
can accommodate the data of the $B^0\to K^\mp\pi^\pm$ direct CP
asymmetry. Our study reconciles the opposite conclusions on the real
or complex penguin annihilation amplitude drawn in the
soft-collinear effective theory and in the perturbative QCD approach
based on $k_T$ factorization theorem.

\end{abstract}


\maketitle

The effect of scalar penguin annihilation on charmless nonleptonic
$B$ meson decays has attracted intensive attention. This
power-suppressed contribution is chirally enhanced, i.e.,
proportional to $\mu_P/m_b$ in $B\to PP$ decays, where $\mu_P$ is
the chiral scale associated with the pseudoscalar meson $P$ and
$m_b$ the $b$ quark mass. Since it involves endpoint singularities,
it was parameterized as a free parameter
$X_A=\ln(m_b/\Lambda)[1+\rho_A\exp(i\phi_A)]$ in QCD-improved
factorization (QCDF) \cite{BBNS}, with $\Lambda$ being a hadronic
scale, and $\rho_A$ and $\phi_A$ varied arbitrarily within some
artificially specified ranges. In order to fit data such as the
$B^0\to K^\mp\pi^\pm$ direct CP asymmetry $A_{\rm CP}(B^0\to
K^\mp\pi^\pm)$, $\phi_A$ must take a sizable value. On the other
hand, the contribution from scalar penguin annihilation has been
found to be almost imaginary in the perturbative QCD (PQCD) approach
based on $k_T$ factorization theorem \cite{KLS,LUY}, and the
resultant strong phase leads to a prediction consistent with the
measured $A_{\rm CP}(B^0\to K^\mp\pi^\pm)$. The annihilation
amplitude was not considered in the leading-power formalism of
soft-collinear effective theory (SCET) \cite{BPS,BPRS,Chay:2003ju}.
Instead, a nonperturbative complex charming penguin was introduced
to accommodate the data of $A_{\rm CP}(B^0\to K^\pm\pi^\mp)$. In the
recent SCET formalism with the zero-bin subtraction \cite{MS06}, the
annihilation contribution becomes factorizable, and has been
concluded to be almost real \cite{ALRS06}.

The motivation of this paper is to reconcile the opposite
theoretical observations on the almost imaginary or almost real
penguin annihilation derived in PQCD and in SCET.  We shall first
point out that the comparison of the measured $A_{\rm CP}(B^\pm\to
K^\pm\pi^0)$ and $A_{\rm CP}(B^\pm\to K^\pm\rho^0)$ indicates an
imaginary penguin annihilation amplitude \cite{LM06,Li07}: The
$B^\pm\to K^\pm\pi^0$ ($B^\pm\to K^\pm\rho^0$) decays involve a
$B\to P$ ($B\to V$) transition, so the penguin emission amplitude is
proportional to the constructive (destructive) combination of the
Wilson coefficients $a_4+(-)2(\mu_K/m_b)a_6$, $\mu_K$ being the
chiral scale associated with the kaon. The annihilation effect is
then less influential in the former than in the latter. If the
penguin annihilation is real, both decays will exhibit small direct
CP asymmetries, i.e., $A_{\rm CP}(B^\pm\to K^\pm\pi^0)\approx A_{\rm
CP}(B^\pm\to K^\pm\rho^0)\approx 0$. If imaginary, it will cause a
larger $A_{\rm CP}(B^\pm\to K^\pm\rho^0)$. The current data $A_{\rm
CP}(B^\pm\to K^\pm\pi^0)=0.050\pm 0.025$ and $A_{\rm CP}(B^\pm\to
K^\pm\rho^0)=0.31^{+0.11}_{-0.10}$ \cite{HFAG} favor an imaginary
penguin annihilation.

We emphasize that strong phases, generated by subleading
corrections, are the leading effect for direct CP asymmetries of $B$
meson decays. For example, the prediction for the direct CP
asymmetry $A_{\rm CP}(B^\pm\to K^\pm\pi^0)$ is sensitive to the
strong phase of the ratio $C/T$ \cite{Charng2,LMS05}, where $C$
($T$) is the color-suppressed (color-allowed) tree amplitude, though
the branching ratio $B(B^\pm\to K^\pm\pi^0)$ is not. Assuming this
ratio to be real as in the leading-power SCET \cite{BPRS}, it is
difficult to explain the data. Therefore, the study of strong phases
requires a careful treatment of subleading corrections. It will be
explained that the different penguin annihilation effects observed
in PQCD and SCET arise from whether parton transverse momenta $k_T$
and other intrinsic mass scales in particle propagators are expanded
or not. If these small scales are neglected or expanded, the
internal particles in an annihilation amplitude are on their mass
shell only at the endpoints of parton momentum fractions, where
hadron distribution amplitudes usually vanish, or the zero-bin
subtraction applies. An annihilation amplitude is then real.
Including $k_T$, the on-shell condition of internal particles does
not occur at the endpoints, so that there is a potential to generate
a sizable strong phase. We claim that when $m_b$ approaches
infinity, the on-shell region coincides with the endpoints, and the
same vanishing results for strong phases will be derived,
irrespective of whether the small scales are expanded into a power
series. For the physical value of $m_b$, however, a formally
power-suppressed correction may have a significant numerical effect
on strong phases, and lead to large direct CP asymmetries in $B$
meson decays.

As argued in Ref.~\cite{Li0408}, a parton, carrying a transverse
momentum $k_T$ as small as a hadronic scale $\Lambda$ initially,
accumulates its $k_T$ after emitting infinitely many collinear
gluons. When the parton participates in a hard scattering eventually,
$k_T$ can become as large as the hard scale. Such an accumulation is
described by the Dokshitzer-Gribov-Lipatov-Altarelli-Parisi
evolution \cite{DGLAP} for a parton distribution function in
inclusive processes and by the Sudakov evolution \cite{CS} for a
hadron wave function in exclusive processes. For two-body
nonleptonic $B$ meson decays, $k_T^2$ of internal particles in a
hard kernel reaches the hard scale of $O(m_b\Lambda)$. That is, the
effect resulting from $k_T^2$ is suppressed by a power of
$r=k_T^2/m_b^2\sim O(\Lambda/m_b)$. In SCET, the power counting rule
for $k_T$ is different, which is always treated as being
$O(\Lambda)$, and expanded. However, there exists a scale of  $O(m_b
\Lambda)$ from the hard-collinear modes, which is also suppressed by
$\Lambda /m_b$ compared to $m_b^2$. To verify the above claim, we
shall keep a small scale in particle propagators, which can be
regarded as an averaged parton transverse momentum in PQCD or the
hard-collinear scale in SCET, and examine its effect on the penguin
annihilation in the SCET formalism with the zero-bin subtraction
\cite{MS06}.

Before computing the direct CP asymmetry of the $B^0\to
K^\mp\pi^\pm$ decays, we illustrate why a formally power-suppressed
correction of $O(r)$ could produce a sizable strong phase in an
annihilation amplitude. Expand a kernel of the form
\begin{eqnarray}
\frac{1}{x-r+i\epsilon}= \frac{1}{x+i\epsilon} +O \Bigl( \frac{r}{x}
\Bigr)  \,,\label{pa}
\end{eqnarray}
which appears in a convolution with a meson distribution amplitude.
Eq.~(\ref{pa}) holds in principle as long as the contribution from
the small $x$ region is suppressed by the meson distribution
amplitude, namely, as the main contribution comes from the region
with $r/x \ll 1$. On the other hand, we have the principle-value
prescription without expansion,
\begin{eqnarray}
\frac{1}{x-r+i\epsilon}=P\frac{1}{x-r}-i\pi\delta(x-r) \,.\label{rq}
\end{eqnarray}
Convoluting the kernel with the distribution amplitude
$\phi(x)=6x(1-x)$, the real parts from Eqs.~(\ref{pa}) and
(\ref{rq}) differ by only 15\%. The imaginary part from
Eq.~(\ref{pa}) vanishes, but that from Eq.~(\ref{rq}) reaches half
of the real part for a typical value of $r\sim \Lambda /m_b\sim
0.1$. Obviously, in order that the imaginary part becomes
negligible, i.e., about 5\% of the real part, $r$ must decrease to
0.01 (or $m_b$ increases up to 50 GeV). The lessons we learn from
this simple example are 1) as $x$ has the substantial probability to
be close to $r$, which is small but away from the endpoint, the
expansion in a power series of $r$ breaks down, and an imaginary
piece could develop; 2) the expansion is reliable only for
sufficiently small $r$ such that the contribution from $x\sim r$ is
highly-suppressed like the endpoint one; 3) $r$ is expected to give
a minor (larger) effect on branching ratios (direct CP asymmetries)
of $B$ meson decays.

Let the momenta of the outgoing quark $u$ and antiquark $\bar u$ in
opposite directions be $k_2=(0,yP_2^-,{\bf 0}_T)$ and $k_3=(\bar
xP_3^+,0,{\bf 0}_T)$, respectively, for the decay $\bar B^0\to
K^-\pi^+$, where $P_2$ ($P_3$) is the pion (kaon) momentum and $\bar
x=1-x$. We quote the expression for the penguin annihilation amplitude
in the SCET formalism with the zero-bin subtraction \cite{ALRS06},
\begin{align} \label{Kpi}
  A_{Lann}(K^-\pi^+) &= -\frac{G_F f_B f_{K} f_{\pi} }{\sqrt{2}}
  (\lambda_c^{(s)}+\lambda_u^{(s)}) \frac{4\pi\alpha_s(\mu_h)}{9}
  \nonumber\\
 &\quad \times \bigg\{ \Big(\frac{C_9}{6}-\frac{C_3}{3}\Big) \Big[
  \big\langle \bar x^{-2}\big\rangle^K \big\langle y^{-1} \big\rangle^\pi
 - \big\langle [y(x\bar y-1)]^{-1} \big\rangle^{\pi K} \Big] \nonumber\\
 & \qquad
 -\frac{2\mu_\pi}{3m_b}
  \Big({C_6}\!-\!\frac{C_8}{2}\!+\!\frac{C_5}{3}\!-\!\frac{C_7}{6}\Big) \Big[
  \big\langle y^{-2} \bar y^{-1} \big\rangle_{pp}^{\pi}
  \big(\big\langle \bar x^{-2}\big\rangle^K \!+\! \big\langle \bar
   x^{-1}\big\rangle^K \big)
 \nonumber\\
 &\qquad
  - \frac{2\mu_\pi}{3m_b}
\Big(\frac{C_5}{3}\!-\!\frac{C_7}{6}\Big)
   \big\langle [(1-x \bar y) \bar x
   y^2]^{-1}\big\rangle^{\pi K}_{pp}
 +\frac{2\mu_K}{3m_b} \Big(\frac{C_5}{3}\!-\!\frac{C_7}{6}\Big)
   \big\langle [(1-x\bar y) \bar x^2 y]^{-1}
  \big\rangle_{pp}^{K\pi}
 \nonumber\\
& \qquad
  -\frac{2\mu_K}{3m_b}
   \Big({C_6}\!-\!\frac{C_8}{2}\!+\!\frac{C_5}{3}\!-\!\frac{C_7}{6}\Big)
  \Big[ \big(\big\langle y^{-2} \big\rangle^\pi+ \big\langle y^{-1}
  \big\rangle^\pi \big)
 \big\langle x^{-1}\bar x^{-2} \big\rangle^K_{pp} \Big]
  \bigg\}\;,
\end{align}
where $G_F$ is the Fermi constant, $f_{B,K,\pi}$ the meson decay
constants, $\lambda_{u,c}^{(s)}$ the products of the
Cabibbo-Kobayashi-Maskawa (CKM) matrix elements, $\mu_h\sim m_b$ the
hard scale, $C_i$ the Wilson coefficients, and $\mu_{\pi}$ the
chiral scale associated with the pion. The logarithmic terms
$\ln\mu_\pm$ in Ref.~\cite{ALRS06} have been dropped since they
are cancelled by the corresponding logarithms in the convolutions.
Because of the large theoretical uncertainty shown below, the constant
$\kappa$ resulting from the above logarithmic cancellation will be
neglected \cite{ALRS06} . The three-parton twist-3 contribution to the
penguin annihilation, being numerically smaller by one order of
magnitude than Eq.~(\ref{Kpi}) \cite{Arnesen:2006dc}, is not included.

\begin{figure}[t]
\begin{center}
\includegraphics{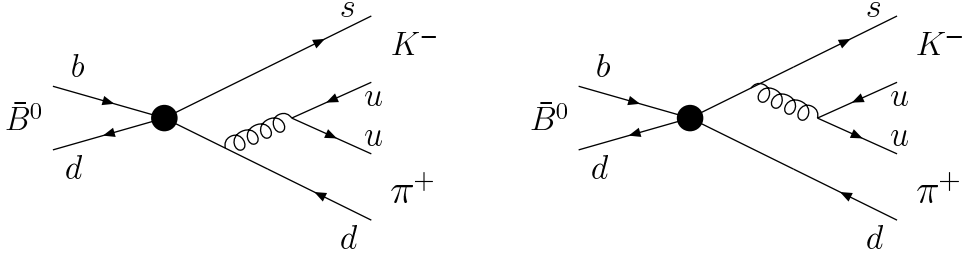}
\caption{
``Factorizable'' annihilation diagrams in the $\bar B^0\to K^-\pi^+$
 decay, where the black dots denote a scalar-penguin operator in the
 effective weak Hamiltonian.}
\label{fig:annihilation}
\end{center}
\end{figure}
Motivated by the illustration based on Eqs.~(\ref{pa}) and
(\ref{rq}), we introduce a small constant $r$ into internal quark
propagators involved in the factorizable piece of Eq.~(\ref{Kpi}),
corresponding to Fig.~\ref{fig:annihilation}. Inserting $r$ into
internal gluon propagators generates a strong phase down by a factor
three. The strong phase from the nonfactorizable annihilation
amplitude is smaller by two orders of magnitude. We stress that
adding $r$ in the aforementioned way causes a double counting of the
contributions from higher-order operators in SCET, and should be
regarded as only a test of our claim. Applying the principle-value
prescription, we obtain the extra imaginary pieces via the following
substitutions,
\begin{eqnarray}
\langle \bar x^{-2}\rangle^M&\to& \langle \bar x^{-2}\rangle^M+i{\rm
Im} \langle \bar x^{-2} \rangle^M,\nonumber\\
& &{\rm Im}\langle \bar x^{-2}
\rangle^M\,=\,-\pi\int_0^1dx\frac{\phi_M(x)+\bar
x\phi_{M}^\prime(1)}{\bar x}\delta\left(\bar
x-r\right)\,,\label{m}\\
\langle y^{-2}\bar y^{-1}\rangle^M_{pp}&\to&\langle y^{-2}\bar
y^{-1}\rangle^M_{pp}+i{\rm Im} \langle y^{-2}\bar
y^{-1} \rangle^M_{pp}\;,\nonumber\\
& &
 {\rm Im}\langle y^{-2}\bar y^{-1} \rangle^M_{pp}\,=\,-\pi\int_0^1dy
\left[\frac{\phi_{pp}^M(y)}{y(1-y)}
-\frac{y\phi_{pp}^{M\prime}(0)}{y}\right]
\delta\left(y-r\right)\,.\label{2nd}
\end{eqnarray}
Employing the parameterizations for the leading-twist distribution
amplitude $\phi_M(x)$ and for the two-parton twist-3 distribution
amplitudes $\phi_{pp}^M(x)$ \cite{ALRS06,Arnesen:2006dc}
\begin{eqnarray}
& &\phi_M(x)\ =\
  6x(1-x)
  \left[1+ a_1^M (6x-3) + 6 a_2^M ( 1- 5 x+ 5x^2)
  - 10 a_3^M ( 1 - 9x + 21 x^2 - 14x^3)
\nonumber\right.\\
& &\left.\hspace{35mm}
  + 15 a_4^M ( 1 - 14x + 56 x^2 -84x^3+42x^4)+\cdots
\right],\nonumber\\
& &\phi_{pp}^M(x) \ =\ 6x(1-x) \left[1+ a_{1pp}^M (6x-3) + 6
a_{2pp}^M ( 1- 5 x+ 5x^2)+\cdots\right],\label{2p}
\end{eqnarray}
with $M=\pi$, $K$, it is easy to find that both Eqs.~(\ref{m}) and
(\ref{2nd}) are proportional to $r$ as expected.

The importance of the penguin annihilation contribution relative to
the full penguin one has been estimated in SCET \cite{ALRS06}, and
found to be about 10\% with large uncertainty in the $B^0\to
K^\mp\pi^\pm$ decays. The full penguin contribution does not come
from an explicit evaluation in the same SCET framework, but from a
fitting to the $B\to K\pi$ data. We can certainly follow this
approach. However, the factorization formulas for the emission
amplitudes have been available in Ref.~\cite{MS06}, so they will be
adopted in the numerical analysis below. The feature of generating
strong phases does not depend on how we estimate the emission
amplitudes. Besides, we shall not include the free parameters
associated with the long-distance charming penguin, which is not
factorizable in SCET. As demonstrated later, a decay amplitude under
the zero-bin subtraction is very sensitive to higher Gegenbauer
moments $a_n^M$ and $a_{npp}^M$ in Eq.~(\ref{2p}) \cite{Feld},
which are mostly unknown. Hence, we shall determine these moments
by fitting the SCET formulas to data of branching ratios, which are
then used to predict direct CP asymmetries. If a strong phase from
the source considered here is sizable, the whole CP asymmetry cannot
be attributed to the nonperturbative charming penguin alone.

At lowest order in $\alpha_s(m_b)$ with the Wilson coefficients
$T^{(+)}=1$ and $C_J^{(+)}=1$ in SCET$_{\rm I}$ \cite{BPS}, the
$B\to\pi$ transition form factor is decomposed into
\begin{eqnarray}
f_+(E) = \zeta^{B\pi}(E) +\zeta_J^{B\pi}(E)\,.\label{btopi}
\end{eqnarray}
The second term is factorizable, written as
\begin{align} \label{f+tree}
  \zeta_J^{B\pi}(E) =  \frac{f_B f_\pi m_B}{4E^2}\
    \frac{4\pi\alpha_s(\mu_i)}{9} \Big( \frac{2E}{m_B}+\frac{2E}{m_b} -1 \Big)
    \int_0^1\!\!\! dy \: \frac{\phi_\pi(y)}{y}\:
    \int_0^\infty\!\!\! dk^+\: \frac{\phi_B^+(k^+)}{k^+} \;,
\end{align}
where $\mu_i\sim\sqrt{m_b\Lambda}$ is the intermediate scale, and $k^+$
the momentum of the spectator quark in the $B$ meson. For charmless
two-body nonleptonic $B$ meson decays, we take the pion energy
$E=m_B/2$, $m_B$ being the $B$ meson mass.
The first term also becomes factorizable after
implementing the zero-bin subtraction for the endpoint singularity
\cite{MS06},
\begin{align} \label{zetafinite}
  \zeta^{B\pi}(E) &= \frac{f_B f_\pi m_B}{4 E^2} \:
  \frac{4\pi\alpha_s(\mu_i)}{9} \!
  \int_0^1\!\!  dy  \int_0^\infty\!\! dk^+
  \left\{
  \frac{ (1\!+\!y)\phi_\pi(y)}{(y^2)_{\mbox{\o}}  }
  \: \frac{\phi_B^-(k^+)}{ (k^+)_{\mbox{\o}} }
+ \mu_\pi \,
   \frac{(\phi_\pi^p\!+\!\frac16\,
     \phi_\pi^{\sigma\prime})(y)}{(y^2)_{\mbox{\o}} }
   \: \frac{\phi_B^+(k^+)}{ (k^{+2})_{\mbox{\o}} }\right\},
\end{align}
where only the terms from the two-parton pion distribution
amplitudes are kept. The relation among $\phi_\pi^p$,
$\phi_\pi^\sigma$ and $\phi_{pp}^\pi$ can be found in
Ref.~\cite{ALRS06}. The formulas for the $B\to K$ form factor in
SCET are similar. We multiply Eq.~(\ref{btopi}) by the appropriate
CKM matrix elements and Wilson coefficients, including a part of
next-to-leading-order corrections~\cite{Jain:2007dy}, to obtain the
emission contributions from both the tree and penguin operators. The
Wilson coefficient $a_6$ was neglected in the previous SCET
analysis, since the associated penguin contribution is
power-suppressed. However, it is enhanced by the chiral scale, and
numerically crucial. Furthermore, the power-suppressed annihilation
has been formulated into SCET, so there is no reason for ignoring
$a_6$ \cite{Jain:2007dy}.

The zero-bin subtraction for the logarithmic endpoint singularity
associated with the pion distribution amplitude $\phi_\pi$ in the
first term of Eq.~(\ref{zetafinite}) is referred to
Ref.~\cite{MS06}, where the term proportional to $y$ in $(1+y)$ does
not require subtraction. We also need the zero-bin subtraction for
the linear endpoint singularity present in the second term of
Eq.~(\ref{zetafinite}) \cite{IS}:
\begin{eqnarray} \label{linear}
 \int_0^1\!\! dy\:
  \frac{ \phi_\pi^p(y) }{(y^2)_{\mbox{\o}} } &\equiv&
  \int_0^1\!\! dy\,
  \frac{ \phi_\pi^p(y) -\phi_\pi^p(0) - y \phi^{p\prime}_\pi(0)}{y^2}\
-\int_1^\infty\!\! dy\,y^\epsilon (y-1)^\epsilon
  \frac{\phi_\pi^p(0) + y \phi^{p\prime}_\pi(0)}{y^2}
  \left(\frac{p^-}{\mu_-}\right)^{2\epsilon}\
  \nonumber\\
&=&\int_0^1\!\! dy\,
  \frac{ \phi_\pi^p(y) -\phi_\pi^p(0) - y
  \phi^{p\prime}_\pi(0)}{y^2}\
  -\phi_\pi^p(0) +
\ln\Big( \frac{\bar n\cdot P_2}{\mu_-}\Big)\: \phi^{p\prime}_\pi(0)
\,,
\end{eqnarray}
where $\overline{n}\cdot P_2 = 2E$. The subtraction
associated with the derivative of the two-parton twist-3 pion
distribution amplitude, $\phi_\pi^{\sigma\prime}$, is similar.

We consider the models for the $B$ meson distribution amplitudes
$\phi_B^\pm$ proposed by Kodaira et al. (KKQT) \cite{KKQT} and by
Grozin and Neubert (GN) \cite{GN}. The associated
zero-bin subtraction is defined by
\begin{eqnarray} \label{Abf}
 \int_0^\infty\!\! dk^+\:
  \frac{ \phi^-_B(k^+) }{(k^+)_{\mbox{\o}} }
   &\equiv& \int_0^\infty\!\! dk^+\,
  \frac{ \phi^-_B(k^+)}{k^+}-\int_0^{\bar\Lambda}\!\! dk^+\,
  \frac{ \phi^-_B(0)}{k^+}\
   + \ln\Big(\frac{n\cdot v\bar\Lambda}{\mu_+}\Big)
   \phi^-_B(0)  \,, \\
  &=&\left\{ \begin{array}{lc}
\displaystyle -\frac{1}{\bar\Lambda}(1-\ln 2)+\ln\Big(\frac{n\cdot
  v\bar\Lambda}{\mu_+}\Big)
   \phi^-_B(0)\,, &\mathrm{for} \ \mathrm{KKQT} \\
\displaystyle   -\frac{1}{\omega_0}\left(\gamma_E+
   \ln\frac{\bar\Lambda}{\omega_0}\right)+\ln\Big(\frac{n\cdot
  v\bar\Lambda}{\mu_+}\Big)
   \phi^-_B(0)\,, &\mathrm{for} \ \mathrm{GN}
\end{array}
\right. \nonumber
\end{eqnarray}
\begin{eqnarray}
 \int_0^\infty\!\! dk^+\:
  \frac{ \phi_B^+(k^+) }{(k^{+2})_{\mbox{\o}} } &\equiv&
  \int_0^\infty\!\! dk^+\,
  \frac{ \phi_B^+(k^+)}{k^{+2}}-\int_0^{\bar\Lambda}\!\! dk^+\,
  \frac{\phi^{+\prime}_B(0)}{k^{+}}\
   + \ln\Big(\frac{n\cdot v\bar\Lambda}{\mu_+}\Big)
   \phi^{+\prime}_B(0)\,,\label{phi+}\\
&=& \left\{ \begin{array}{lc} \displaystyle
\frac{1}{2\bar\Lambda^2}\ln 2+\ln\Big(\frac{n\cdot
v\bar\Lambda}{\mu_+}\Big)
   \phi^{+\prime}_B(0)\,,&\mathrm{for}\ \mathrm{KKQT} \\
\displaystyle -\frac{1}{\omega_0^2}\left(\gamma_E+
   \ln\frac{\bar\Lambda}{\omega_0}\right)+\ln\Big(\frac{n\cdot
  v\bar\Lambda}{\mu_+}\Big)
   \phi^{+\prime}_B(0)\,, &\mathrm{for} \ \mathrm{GN}
        \end{array}
\right. \nonumber
\end{eqnarray}
with the parameter relation $\omega_0=2\bar\Lambda/3$, $\bar\Lambda$
being the $B$ meson and $b$ quark mass difference. In the above
expressions $n$ is a light-like vector along the Wilson line in the
definition for the $B$ meson distribution amplitudes, and $v$ is the
$B$ meson velocity. The terms containing $\ln\mu_\pm$ in
Eqs.~(\ref{linear})-(\ref{phi+}) are also dropped.

For the numerical analysis, we assume the Gegenbauer moments of the
pion and kaon distribution amplitudes, $a_1^\pi = 0.0$,
$a_{1}^K=-0.05$ consistent with the results in
Ref.~\cite{Braun:2004vf,Ball:2006wn}, $a_{2}^K=a_{2}^\pi=0.2$
\cite{Ball:2006wn,KMM04,BZ05}, $a_3^\pi=0$, $a_{4}^K=a_{4}^\pi$,
$a_{1pp}^\pi = a_{1pp}^K = 0.0$, and $a_{2pp}^K=a_{2pp}^\pi$, among
which $a_4^M$ and $a_{2pp}^M$ are most uncertain. To simplify the
formulas, we do not consider the Gegenbauer moment $a_3^K$ for the
twist-2 kaon distribution amplitude. That is, we keep one most
uncertain parameter from each of $\phi_M$ and $\phi_{pp}^M$, whose
variation is sufficient for our purpose. The hard and intermediate
scales are fixed at $\mu_h=m_b$ and $\mu_i=\sqrt{\bar\Lambda m_b}$,
respectively, with $\bar\Lambda=0.55$ GeV and $m_b=m_b^{1S}= 4.7$
GeV. Other relevant heavy-quark masses are taken to be
$m_c=m_c^{1S}=1.4$ GeV and $\overline m_b=\overline
m_b^{\overline{\rm MS}}(\overline m_b)=4.2$ GeV. We obtain the
chiral scales $\mu_\pi(\mu_h) = 2.4$ GeV, $\mu_K(\mu_h) = 3.0$ GeV,
$\mu_\pi(\mu_i) = 1.8$ GeV, and $\mu_K(\mu_i) = 2.3$ GeV from the
two-loop running for the strong coupling constant with
$\alpha_s(M_Z=91.1876\,{\rm GeV})=0.118$ and for the light-quark
masses with $m_{u,d}(2\,{\rm GeV})=5$ MeV and $m_s(2\,{\rm GeV})=95$
MeV. We take the Wilson coefficients for four-fermion operators
evaluated at $\mu_h=m_b$ and at next-to-leading-logarithmic level:
$C_{1}= 1.078$, $C_{2}= -0.177$, $C_{3}= 0.014$, $C_{4}= -0.034$,
$C_{5}= 0.009$, $C_{6}= -0.040$, $C_{7}= 0.7 \times 10^{-4}$,
$C_{8}= 4.5 \times 10^{-4}$, $C_{9}= -9.9 \times 10^{-3}$, and
$C_{10}= 1.8 \times 10^{-3}$. Those for dipole operators at
leading-logarithmic level are $C_{7\gamma}= -0.314$ and $C_{8G}=
-0.149$ \cite{Buchalla:1995vs}. We also take the Fermi constant
$G_F=1.16639\times 10^{-5}$ GeV$^{-2}$, the decay constants $f_B  =
0.22$ GeV, $f_K = 0.16$ GeV, and $f_\pi =0.131$ GeV, the meson
masses $m_B = 5.28$ GeV, $m_K = 0.497$ GeV, and $m_\pi = 0.14$ GeV,
the $B$ meson lifetime $\tau_B^0=1.530\times 10^{-12}$ sec, and the
CKM matrix elements $V_{us}=0.2257$, $V_{ub}=(4.2\times
10^{-3})\exp(-i\phi_3)$, $V_{cs}=0.957$, and $V_{cb}=0.0416$ with
the weak phase $\phi_3=74^\circ$ \cite{PDG}.

Adopting the above parameters, the two pieces $\zeta^{B\pi}$ and
$\zeta_J^{B\pi}$ of the $B\to\pi$ form factor are written as
\begin{eqnarray}
\zeta^{B\pi} &=& \left\{
\begin{array}{ll}
\displaystyle 0.01 + 0.75\, a_2^\pi + 2.57\, a_4^\pi + 0.43\,
a_{2pp}^\pi\,,
& \;\;\mbox{\rm for KKQT}\label{kk}\\
\displaystyle 0.09 + 0.65\, a_2^\pi + 2.23\, a_4^\pi - 2.73\,
a_{2pp}^\pi\,,
& \;\;\mbox{\rm for GN}\label{gn}\\
\end{array}
\right.
\\
\zeta_J^{B\pi} &=& \left\{
\begin{array}{ll}
\displaystyle 0.016(1.0+a_2^\pi+a_4^\pi)\,,
& \;\;\mbox{\rm for KKQT}\\
\displaystyle 0.024(1.0+a_2^\pi+a_4^\pi)\,,
& \;\;\mbox{\rm for GN}.\\
\end{array}
\right.
\end{eqnarray}
Note that the coefficients in Eq.~(\ref{kk}) grow quadratically with
the order $n$ of the Gegenbauer moments $a_n^\pi$ \cite{Feld}. This
sensitivity is attributed to the increasing slope of the higher
Gegenbauer polynomials at the endpoints of the momentum fraction
$x$. The sign flip of the $a_{2pp}^\pi$ terms indicates that
$\zeta^{B\pi}$ also depends strongly on the models of the $B$ meson
distribution amplitudes in SCET. We mention that the PQCD approach
does not suffer such sensitivity, because the endpoint singularity
is smeared by including parton transverse momenta $k_T$, whose order
of magnitude is governed by the Sudakov factor.

The strong dependence on the higher Gegenbauer moments also appears
in the penguin annihilation amplitude,
\begin{eqnarray}
10^{4}\hat P^{\rm ann}_{K\pi} &\equiv&
- 10^{4}\frac{\sqrt{2}}{G_F
m_B^2}\frac{A_{Lann}(K^-\pi^+)}{(1 {\rm GeV})}\;,\nonumber\\
&=&2.76 ( 0.07 + a_4^\pi)(1.20 + a_4^\pi) + a_{2pp}^\pi (27.0 +
413.1\, a_4^\pi)
- i \pi\, r\, a_{2pp}^\pi ( 53.2 + 1747\, a_4^\pi )
\,,\label{anni}
\end{eqnarray}
with a significant growth of the coefficients of $a_4^\pi$. The
imaginary contribution is proportional to the second moment
$a_{2pp}^\pi$. In fact, it could depend on the zeroth moment, i.e.,
the normalization of $\phi_{pp}^M$, if the denominator $1-y$ is not
replaced by 1 in the subtraction term in Eq.~(\ref{2nd}). The
denominators $1-y$ and 1 correspond to different zero-bin
subtraction schemes.

Note that the size of the imaginary part depends on the amount of the
subtracted contribution, i.e., on zero-bin subtraction schemes, since it
is generated at $\bar x\sim\Lambda/m_b$ or $y\sim\Lambda/m_b$ as shown
in Eqs.~(\ref{m}) and (\ref{2nd}). The dependence on subtraction schemes
also exists in all other definitions like
Eqs.~(\ref{linear})-(\ref{phi+}), which will not be discussed in this
work.

For the range of $a_4^\pi$, the crude bound $a_4^\pi\geq -0.07$ has
been determined in Ref.~\cite{BZ05}. The analysis based on the data
of the pion transition form factor suggests $a_4^\pi\approx
-0.05$ in Ref.~\cite{SSA} and the constraint
$a_2^\pi+a_4^\pi=-0.03\pm 0.14$ in Ref.~\cite{BSM}, both of which
prefer a negative value of $a_4^\pi$ (considering $a_2^\pi\approx
0.2$). The range of $a_{2pp}^\pi$ is basically undetermined. We
shall regard these two parameters as being free, and fix them by the
strategy stated before: Adjust $a_4^\pi$ and $a_{2pp}^\pi$, such
that the $B\to\pi$ form factor has the value around $f_+=0.24\pm
0.05$ \cite{LMS05}, and the $B^0\to K^\mp\pi^\pm$ decays have the
branching ratio close to the data $B(B^0\to K^\mp\pi^\pm)=(19.4\pm
0.6)\times 10^{-6}$ \cite{HFAG}.
Because the last two terms in $\zeta^{B\pi}$ for the KKQT model are
of the same sign, and the coefficient of $a_4^\pi$ is large, the
constraint from the form factor value leads to a smaller $a_4^\pi$.
Eq.~(\ref{anni}) then implies that the coefficient of $r$,
i.e., the imaginary part of the annihilation amplitude, is smaller,
and that the strong phase is less sensitive to the variation of $r$.
On the contrary, the last two terms in $\zeta^{B\pi}$ for the GN
model have the coefficients with the same order of magnitude, but in
opposite signs. Hence, $a_4^\pi$ (and also $a_{2pp}^\pi$) is larger,
and the strong phase is more sensitive to the variation of $r$ in
this case.

Employing the KKQT model for the $B$ meson distribution amplitudes,
we obtain $a_4^\pi \approx 0.01$ and $a_{2pp}^\pi\approx 0.23$,
corresponding to which the $B\to\pi$ form factor, the $B^0\to
K^\mp\pi^\pm$ branching ratio, and the predicted direct CP asymmetry
are given by
\begin{eqnarray}
\zeta^{B\pi} &=& 0.29
\,,\nonumber\\
\zeta_J^{B\pi} &=& 0.02
\,,\nonumber\\
B(B^0\to K^\mp\pi^\pm) &=& \left\{
\begin{array}{ll}
20.5 \times 10^{-6}
& \mbox{\rm for } r=0.0\\
20.0 \times 10^{-6}
& \mbox{\rm for } r=0.1\\
19.8 \times 10^{-6}
& \mbox{\rm for } r=0.2\,,\\
\end{array}
\right.
\nonumber\\
A_{\rm CP}(B^0\to K^\mp\pi^\pm) &=& \left\{
\begin{array}{ll}
0.08
& \mbox{\rm for } r=0.0\\
0.05
& \mbox{\rm for } r=0.1\\
0.02
& \mbox{\rm for } r=0.2\,.\\
\end{array}
\right.\label{KK1}
\end{eqnarray}
We do not attempt a fine tuning here, but accept the values of
$a_4^\pi$ and $a_{2pp}^\pi$ as solutions, when they produce the
$B\to\pi$ form factor and the $B^0\to K^\mp\pi^\pm$ branching ratio
close to the designated ranges. The results shift with the slight
variation of $a_4^\pi$ and $a_{2pp}^\pi$, but the behavior for
different $r$ in Eq.~(\ref{KK1}) has the same pattern. In principle,
$\zeta^{B\pi}$ and $\zeta^{B\pi}_J$ have the same scaling law in
$\alpha_s$ and in $1/m_b$ \cite{TLS,Bauer:2005kd}. The numerical
hierarchy $\zeta^{B\pi}\gg \zeta^{B\pi}_J$ in Eq.~(\ref{KK1}),
consistent with the PQCD results \cite{TLS}, may be altered in
different zero-bin subtraction schemes. It is obvious that the power
correction associated with $r$ has a negligible effect on the
branching ratio. However, the power correction generates a strong
phase: $A_{\rm CP}(B^0\to K^\mp\pi^\pm)$ decreases by 40\% from
$r=0$ to $r=0.1$. Since the imaginary part is proportional to $r$,
it is difficult to accommodate the data $A_{\rm CP}(B^0\to
K^\mp\pi^\pm)=-0.097\pm 0.012$ \cite{HFAG} with a reasonable value
of the power-suppressed $r$ using the KKQT model.

For the GN model, we find two sets of solutions corresponding to
$a_4^\pi\approx 0.18$ and $a_{2pp}^\pi\approx 0.15$,
\begin{eqnarray}
\zeta^{B\pi} &=& 0.21
\,,\nonumber\\
\zeta_J^{B\pi} &=& 0.03
\,,\nonumber\\
B(B^0\to K^\mp\pi^\pm) &=& \left\{
\begin{array}{ll}
20.1 \times 10^{-6}
& \mbox{\rm for } r=0.0\\
20.4 \times 10^{-6}
& \mbox{\rm for } r=0.1\\
25.1 \times 10^{-6}
& \mbox{\rm for } r=0.2\,,\\
\end{array}
\right.
\nonumber\\
A_{\rm CP}(B^0\to K^\mp\pi^\pm) &=& \left\{
\begin{array}{ll}
0.06
& \mbox{\rm for } r=0.0\\
-0.06
& \mbox{\rm for } r=0.1\\
-0.14
& \mbox{\rm for } r=0.2\,,\\
\end{array}
\right.\label{gn1}
\end{eqnarray}
and to $a_4^\pi\approx -0.22$ and $a_{2pp}^\pi\approx -0.20$,
\begin{eqnarray}
\zeta^{B\pi} &=& 0.28
\,,\nonumber\\
\zeta_J^{B\pi} &=& 0.02
\,,\nonumber\\
B(B^0\to K^\mp\pi^\pm) &=& \left\{
\begin{array}{ll}
18.6 \times 10^{-6}
& \mbox{\rm for } r=0.0\\
19.4 \times 10^{-6}
& \mbox{\rm for } r=0.1\\
26.5 \times 10^{-6}
& \mbox{\rm for } r=0.2\,,\\
\end{array}
\right.
\nonumber\\
A_{\rm CP}(B^0\to K^\mp\pi^\pm) &=& \left\{
\begin{array}{ll}
0.08
& \mbox{\rm for } r=0.0\\
-0.10
& \mbox{\rm for } r=0.1\\
-0.20
& \mbox{\rm for } r=0.2\,.\\
\end{array}
\right.\label{gn2}
\end{eqnarray}
The existence of the two sets of solutions with opposite signs is
understandable. Because the term proportional to $a_4^\pi$ in the
imaginary part of Eq.~(\ref{anni}) dominates over the constant term
as $|a_4^\pi|$ reaches about 0.2, the product $a_{2pp}^\pi a_4^\pi$
matters, and $a_4^\pi$ and $a_{2pp}^\pi$ can flip sign
simultaneously.

As indicated by Eqs.~(\ref{gn1}) and (\ref{gn2}), the branching
ratio is stable, while the strong phase is very sensitive to the
variation of $r$, so that we easily accommodate the data of $A_{\rm
CP}(B^0\to K^\mp\pi^\pm)$ with a typical value of $r=0.1\sim 0.15$.
The predicted $A_{\rm CP}(B^0\to K^\mp\pi^\pm)$ for $r=0$, i.e.,
real penguin annihilation ($r=0.1$, i.e., complex penguin
annihilation) is close to that from QCDF in the default scenario
\cite{BN} (PQCD \cite{KLS,LMS05}). Therefore, the strong phases
resulting from the power-suppressed source in the penguin
annihilation could be numerically crucial for the estimation of
direct CP asymmetries. We then understand the opposite conclusions
on the effect of the penguin annihilation drawn in SCET and in PQCD:
The almost real annihilation amplitude in the former and the almost
imaginary annihilation amplitude in the latter are attributed to the
different treatments of the formally power-suppressed terms at the
physical $b$ quark mass. Note that the solutions of $a_4^\pi$ and
$a_{2pp}^\pi$ in Eqs.~(\ref{KK1})-(\ref{gn2}) will be changed, if
higher Gegenbauer moments in Eq.~(\ref{2p}) are taken into account,
which cause even larger variation of the decay amplitudes. However,
the strong dependence of $A_{\rm CP}(B^0\to K^\mp\pi^\pm)$ on $r$
will persist.

SCET provides a systematical expansion in powers of $\Lambda/m_b$,
which is somewhat twisted here by keeping subleading terms in
particle propagators in order to demonstrate a possible mechanism
for generating strong phases. This twist of SCET actually violates
its power counting rules and other aspects. Hence, our analysis does
not imply the breakdown of SCET in its application to $B$ meson
decays, but helps clarifying why there are discrepancies in the
study of direct CP asymmetries from SCET and PQCD. It hints that
more caution is necessary for fixed-power evaluations of direct CP
asymmetries at the physical mass $m_b$. The expansion would be
reliable for decay rates and direct CP asymmetries, if the $b$ quark
mass was 10 times heavier. In that case, the contribution from the
on-shell region of internal particles can be really suppressed by
hadron distribution amplitudes, or excluded by the zero-bin
subtraction. For $m_b\approx 5$ GeV, a novel method might be
demanded.

We have shown that introducing a small scale into denominators of
internal quark propagators can accommodate both the measured
branching ratio and direct CP asymmetry of the $B^0\to K^\mp\pi^\pm$
decays. Keeping a small quantity in denominators without expansion
is equivalent to resummation of the associated corrections to all
powers. It is similar to resummation of part of higher-order
corrections in $\alpha_s$ for many QCD processes. It has been
explained that at least the parton transverse momenta can be
maintained in denominators consistently in $k_T$ factorization
theorem \cite{NL2,NL07}. This treatment is justified by different
power counting rules, which hold in the region of small parton
momenta \cite{NL07}. This alternative power expansion, postulated in
$k_T$ factorization theorem, has led to strong phases in more
agreement with the indication of data in $B$ meson decays.

\vskip 1.0cm We thank Z. Ligeti and I. Stewart for useful
discussions. JC is supported in part by Grant No.
R01-2006-000-10912-0 from the Basic Research Program of the Korea
Science and Engineering Foundation, and by funds provided by the U.
S. Department of Energy (DOE) under cooperative research agreement
DE-FC02-94ER40818. HNL is supported by the National Science Council
of R.O.C. under Grant No. NSC-95-2112-M-050-MY3 and by the National
Center for Theoretical Sciences of R.O.C.. SM is supported by the
U.S. DOE under Grant No. DE-FG02-90ER40542. HNL thanks Korean
Institute for Advanced Studies and Korea University for their
hospitality during his visit, where this work was initiated. SM
acknowledges the Aspen Center for Physics, where a part of this work
was performed.

\end{document}